\documentclass[twocolumn,showpacs,preprintnumbers,amsmath,amssymb]{revtex4}
\usepackage{graphicx}
\usepackage{dcolumn}
\usepackage{bm}
\begin{document}

\preprint{APS/123-QED}

\title{Dirty, Skewed, and Backwards: The Smectic $A$-$C$ Phase Transition in Aerogel}

\author{Leiming Chen and John Toner}

\address{Department of Physics, Institute of Theoretical Science,
University of Oregon, Eugene OR 97403}

\begin{abstract}
We study the smectic AC transition in anisotropic and uniaxial
disordered environments, e.g., aerogel with an external field. We
find very strange behavior of translational correlations: the {\it
low-temperature}, {\it lower-symmetry} Smectic $C$ phase is {\it
less} translationally ordered than the {\it high-temperature},
{\it higher-symmetry} Smectic $A$ phase, with short-ranged and
algebraic translational correlations, respectively. Specifically,
the $A$ and $C$ phase belong to the quasi-long-ranged
translationally ordered `` $XY$ Bragg glass '' and short-ranged
translationally ordered `` $m=1$ Bragg glass '' phase,
respectively. The $AC$ phase transition itself belongs to a new
universality class, whose fixed points and exponents we find in a
$d=5-\epsilon$ expansion.\end{abstract}  \maketitle

\section{introduction}

Of all randomly pinned elastic media \cite{SC, CDW, J, Helium}
liquid crystals in aerogel \cite{LC} exhibit a phenomenon unique
to themselves: anomalous elasticity. That is the scalings of their
elastic energies are changed radically ( specifically, by
non-trivial power laws ).

However, there has been no previous work on phase transitions in
pinned liquid crystal systems.  In this paper we remedy this  by
treating the smectic $A$ to smectic $C$ ( hereafter $AC$ )
transition in an {\it anisotropic}, {\it uniaxial} disordered
environment. Such an environment could be realized, e.g., by
applying an electric or magnetic field to a liquid crystal in {\it
isotropic} aerogel \cite{Karl}, or by stretching the aerogel
uniaxially before absorbing the liquid crystal. We will hereafter
refer to the special uniaxial direction as `` along the applied
field '' or `` the $z$-axis ''.

The $AC$ phase transition separates the two novel glassy phases
discovered in reference \cite{Karl}. The high temperature phase
($T>T_{AC}$) is the glassy analog of the smectic $A$ phase of the
pure problem, in that the layer normals lie, on average, along the
applied field.  This `` random field $XY$ smectic Bragg glass
''($XYBG$) phase is in the universality class of the random field
$XY$ model. The low temperature phase is the glassy analog of the
smectic $C$ phase, in that the layers normals make an angle
$\theta(T)$ with the applied field. The experimentally measurable
`` tilt angle '' $\theta(T)$ is the magnitude of the order
parameter for the transition. This phase is in the universality
class of the `` $m=1$ smectic Bragg glass ''($m=1$) phase studied
in \cite{Karl}.

We call both of these phases `` glassy'' because the random
environment (i.e., the aerogel) destroys long-ranged translational
order in both. The extent of this destruction, however, differs
greatly between the two phases. Strikingly, it is the {\it
low}-temperature, higher-symmetry, $m=1$ phase that has {\it less}
translational order. In the $XYBG$ phase, translational
correlations are `` quasi-long-ranged '', i.e., they decay as
power laws with distance. In the $m=1$ phase, these correlations
are short-ranged. This leads to radically different X-ray
scattering signatures in the two phases which we will now
describe.

In the $XYBG$ phase, the X-ray scattering intensity $I(\vec{q})$
diverges near the smectic Bragg peaks, which occur at
$\vec{q}=nq_0\hat{z}$ for all $n$ integer, where $q_0={{2\pi}\over
a}$, with $a$ the smectic layer spacing. This divergence is a
power-law:
\begin{eqnarray}
I\left(\vec{q}\right)\propto [\left(q_z-nq_0\right)^2+\alpha
q_{\perp}^2] ^{{-3+.55n^2}\over 2}, \label{Intensity}
\end{eqnarray}
where $\alpha$ is a non-universal constant of order 1 and
$q_{\perp}\equiv|\vec{q}-q_z\hat{z}|$. Note only the first 2 peaks
($n=1$ and $n=2$) actually diverge. In contrast, in the `` glassy
$C$ '' or `` $m=1$ $BG$ '' phase, the peaks in the X-ray
scattering intensity are broad, with $I(\vec{q})$ finite for all
$\vec{q}$.

As $T\to T_{AC}$ from above (i.e., on the $A$ side), the sharp
peaks disappear in an unusual way. The peaks look broad for
$\vec{q}$ 's sufficiently far from the Bragg peak position
$nq_0\hat{z}$, while for $\vec{q}$ 's sufficiently close to the
peak, they diverge per eqn.(\ref{Intensity}). `` Sufficiently
close '' means that {\it both} $|\vec{q}_{\perp}|\ll \delta
q_{\perp}^c(n,T)$, {\it and} $|q_z-nq_0|\ll {\delta q_z^c(T)}$,
where
\begin{eqnarray}
\delta q^c_{\perp, z}(n, T)\propto\left(\xi_{\perp,
z}^c\right)^{-{n^2\over {3-0.55n^2}}}\label{}
\end{eqnarray}
with $\xi_{\perp, z}^c\sim
\exp(A\left|T-T_{AC}\right|^{-\Omega})$, where $\Omega$ is a
universal exponent calculated below and $A$ is a non-universal
constant. These predictions are illustrated in figures 1 and 2.
The divergence of $\xi_{{\perp}, z}^c$ implies that, as $T\to
T_{AC}^+$, the algebraic `` spikes '' on top of the broad
short-ranged peaks get narrower and less intense, vanishing
completely at $T_{AC}$. Lowering temperature further leads only to
the broad peaks of the Smectic $C$ phase. This entire scenario of
sharp peaks at high temperature and broad peaks at low temperature
is very counterintuitive, and unlike almost every other
translationally ordered system \cite{foot}.
\begin{figure}
 \includegraphics[width=0.40\textwidth,angle=-90]{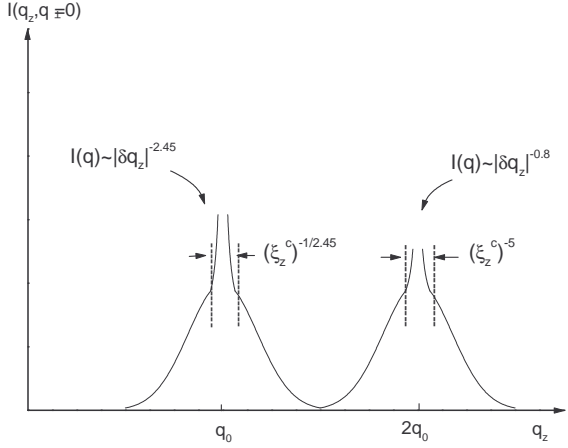}
 \caption{The $q_z$-dependence of the X-ray scattering intensity
 for $q_{\perp}=0$ in the smectic $A$ phase. In the $C$ phase, the
 sharp, power law peaks disappear, leaving only the broad scattering.}
 \end{figure}
\begin{figure}
\includegraphics[width=0.40\textwidth,angle=-90]{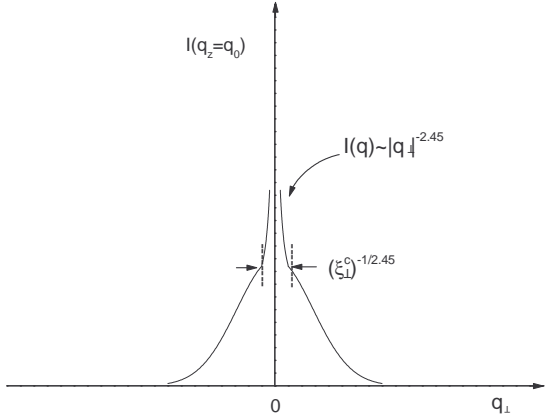}
\caption{The $q_{\perp}$-dependence of the X-ray scattering
intensity for $q_z=q_0$ in the smectic $A$ phase. Again, the sharp
peak vanishes in the $C$ phase.}
\end{figure}

Why is the lower-symmetry $C$ phase {\it less} translationally
ordered than higher-symmetry $A$ phase? In fact, it is precisely
the new broken symmetry of the smectic $C$ phase-that is, the tilt
of the layer normal-that causes this. This is because, while the
energetically preferred layer normal $\hat{N}$ in the smectic $A$
phase is {\it unique}-it must point {\it along} $\hat{z}$-there
are {\it infinitely} {\it many} energetically preferred
orientations of $\hat{N}$ in the $C$ phase: $\hat{N}$ can lie
anywhere on a cone making an angle $\theta(t)$ with $\hat{z}$. As
a result, the elasticity of the Smectic $C$ phase is {\it softer}
than that of the $A$ phase because rotating the layers in such a
way that $\hat{N}$ rotates {\it around} this cone costs no energy.
This {\it exact} symmetry of the elastic energy of the smectic $C$
phase means that the direction perpendicular to the
$\hat{z}$-$<\hat{N}>$ plane (where $<\hat{N}>$ is the mean of
$\hat{N}$; i.e., the direction of spontaneous tilt) becomes ``
soft '', that is, an easy direction for layer displacements to
vary in. Precisely such softness occurs (for different reasons) in
the `` $m=1$ smectic '' studied in \cite{Karl} and, indeed, the
elastic Hamiltonian for the $C$ phase we find is identical to that
studied for ``$m=1$ smectic'' in \cite{Karl}. Thus, we can simply
transcribe the results of \cite{Karl} to this problem.

In particular, positional fluctuations $u$ of the layers about
their optimal (tilted) positions obey
$\overline{<|u(\vec{q})|^2>}=T{C(\vec{q})}^{-1}+(\Delta_s(\vec{q})q_s^2
+\Delta_h(\vec{q})q_h^2+\Delta_zq_z^2){C(\vec{q})}^{-2}$, with
$C(\vec{q})\equiv Bq_z^2+\gamma(\vec{q})q_h^2+K(\vec{q})q_s^4$,
where $<>$ denotes a thermal average, and the overbar denotes an
average over disorder. The anomalous quantities
$\Delta_{s,h}(\vec{q})$, $K(\vec{q})$ and $\gamma(\vec{q})$ obey
\begin{eqnarray}
K, \gamma, \Delta_{s, h}\sim\left\{
\begin{array}{ll}
q_s^{-{\tilde{\eta}_K}, \eta_{\gamma}, -\eta_{s, h}},
&q_s^{\zeta_h}\gg q_{h}, q_s^{\zeta_z}\gg q_{z}
\\
q_{h}^{{-{\tilde{\eta}_K}, \eta_{\gamma}, -\eta_{s, h}}/\zeta_h},
&q_{h}\gg q_s^{\zeta_h}, q_{h}\gg q_z^{\zeta_h\over\zeta_z}
\\
q_{z}^{{-{\tilde{\eta}_K}, \eta_{\gamma}, -\eta_{s, h}}/\zeta_z},
&q_{z}\gg q_s^{\zeta_z}, q_{z}\gg q_{h} ^{\zeta_z\over\zeta_h}
\end{array}\right.
\label{bp}
\end{eqnarray}
where $q_h=Eq_x-Fq_z$ with $E\propto
|T-T_{AC}|^{\nu_{\perp}\left(1-{\eta_K\over 2}\right)}$ and
$F\propto |T-T_{AC}|^{{\nu_{\perp}\over
2}(2+\eta_t-\eta_K-\eta_c)}$, $q_s$ is the component of $\vec{q}$
perpendicular to the $\hat{z}$-$\hat{N}$ plane, $q_x$ is the
component of $\vec{q}$ within the $\hat{z}$-$\hat{N}$ plane
orthogonal to $\hat{z}$ and the critical exponents $\nu_{\perp}$
and $\eta_{t, K, c}$ will be given later. The {\it universal}
exponents in (\ref{bp}) are given by
$\zeta_h=2-({{\eta_{\gamma}+\tilde{\eta}_K}\over 2})$,
$\zeta_z=2-{\tilde{\eta}_K\over 2}$ with the $\eta$'s obeying the
{\it exact} scaling relations $1+\eta_s={\eta_{\gamma}\over
2}+2\tilde{\eta}_{K}$ and
$\eta_h=2+\eta_s-\eta_{\gamma}-\tilde{\eta}_K$. Numerical values
for the $\eta$'s have been obtained \cite{Karl} by comparing
$\epsilon$-expansions based on two different analytical
continuations of this model to higher dimensions, giving
$\tilde{\eta}_K=0.50\pm0.03$, $\eta_{\gamma}=0.26\pm0.12$,
$\eta_s=0.132\pm0.002$ and $\eta_h=1.372\pm0.12$. These
predictions can be tested by light scattering, which measures
$\overline{<N_i(\vec{q})N_j(-\vec{q})>}=q_i^{\perp}q_j^{\perp}\overline
{<|u(\vec{q})|^2>}$. The form given for
$\overline{<|u(\vec{q})|^2>}$ also implies \cite{Karl}
short-ranged translational order and, hence, broad Bragg peaks, as
described earlier.

We have studied the $AC$ transition in an $\epsilon = 5 - d$
expansion, where $d$ is the dimension of space, and find that
there {\it is} a stable fixed point, implying a second-order phase
transition with {\it universal} critical behavior. In particular,
the tilt angle $\theta(T)$ obeys $\theta(T) =
A\left(T_{AC}-T\right)^{\beta}$, where, to leading order in
$\epsilon$, $\beta = {1 \over 2} - {\epsilon \over 10} +
O\left(\epsilon^2\right)$.

The specific heat exponent $\alpha = - {\epsilon \over 10} +
O\left(\epsilon^2\right)$ and the susceptibility exponent $\gamma
= 0$.

The order parameter $\vec{N}_{\perp}$ for this transition is the
projection of the smectic layer normals $\hat{N}$ perpendicular to
the applied field. Above $T_{AC}$, real space correlations of
$\vec{N}_{\perp}$ decay rapidly with distance, with correlation
lengths $\xi_z$ and $\xi_{\perp}$ parallel and perpendicular to
the field respectively. Both diverge as power laws in $( T -
T_{AC} )$: $\xi_{\perp, z} \propto |T - T_{AC}|^{-\nu_{\perp,
z}}$. We find $\nu_{\perp} = {1 \over 2} + {3\epsilon \over 20} +
O\left(\epsilon^2\right)$, $\nu_z = 1 + {3\epsilon \over 10} +
O\left(\epsilon^2\right)$.

We also find that the system exhibits anomalous elasticity right
at $T_{AC}$ as well. Specifically, we find that, right at
$T_{AC}$, the smectic layer bend modulus $K$ {\it vanishes} as
$\vec{q} \rightarrow 0$ according to the scaling laws
\begin{eqnarray}
K\left(\vec{q}\right) = q_{\perp}^{-\eta_K} f_K \left({q_z  \over
q_{\perp}^{\zeta} }\right) \sim\left\{
\begin{array} {ll}
q_{\perp}^{-\eta_K}, &q_z \ll q_{\perp}^{\zeta}\\
q_{z}^{-{\eta_K\over \zeta}}, &q_z \gg q_{\perp}^{\zeta}\\
\end{array}\right.
\label{K}
\end{eqnarray}
where the anisotropy exponent $\zeta = 2 - {\eta _K \over 2}$ and
$\eta_K = C_K \epsilon^2+O(\epsilon^3)$ with $C_K =
{{32\ln\left(4\over 3\right) - 10}\over 225}\cong -.00353$. Note
that $C_K < 0$, which implies that $K$ {\it vanishes} as
$\vec{q}\to 0$. The anisotropy exponent also obeys $\zeta=
{\nu_z\over\nu_{\perp}}$.

The disordering effect of the random aerogel matrix can be
quantified by disorder variances $\Delta_t$ and $\Delta_c$
describing tilt and compressive stresses respectively. These
variances also become anomalous, obeying ($i=t, c$)
\begin{eqnarray}
\Delta_i\left(\vec{q}\right) = q_{\perp}^{-\eta_i} f_{\Delta}
\left({q_z \over q_{\perp}^{\zeta} }\right) \sim \left\{
\begin{array} {ll}
q_{\perp}^{-\eta_i}, &q_z \ll q_{\perp}^{\zeta}\\
q_z^{-{\eta_i\over \zeta}}, &q_z \gg q_{\perp}^{\zeta}\\
\end{array}\right..
\label{Delta}
\end{eqnarray}

Unlike the similar problem of a smectic $A$ in {\it isotropic}
aerogel and no field, the smectic layer compression modulus $B$
remains finite as $\vec{q}\to 0$ at $T_{AC}$.

The exponents $\eta_t$ and $\eta_c$ are given by $\eta_t =
C_{\Delta} \epsilon^2+O(\epsilon^3)$ with $C_{\Delta} =
{{12\ln\left(4\over 3\right) - {1\over 3}}\over 225}\approx
.01386$ and $\eta_c=2-{\epsilon\over5}+O(\epsilon^2)$.

For $T$ bigger than $T_{AC}$ the disorder variance $\Delta_{t, c}
\left(\vec{q}, T\right)$ and layer bend modulus $K \left(\vec{q},
T\right)$ are given by their $T = T_{AC}$ forms equations
(\ref{K}) and (\ref{Delta}) if {\it either} $q_{\perp}\xi_{\perp}
\gg 1$ or $q_z\xi _z\gg 1$.  Otherwise (i.e., if {\it both}
$q_{\perp}\xi_{\perp} \ll 1$ and $q_z\xi _z\ll 1$), $\Delta_{t, c}
\propto \xi^{\eta_{t, c}}_{\perp}\propto\left(T - T_{AC}
\right)^{-\nu_{\perp}\eta_{t, c}}$ and $K
\propto\xi^{\eta_K}_{\perp} \propto \left(T
-T_c\right)^{-\nu_{\perp}\eta_K}$.

The critical exponents obey {\it exact} scaling relations:
\begin{eqnarray}
\alpha &=& 2 - \nu _{\perp} \left(d - 1 + {\eta _K \over 2} - \eta
_t \right) \label{alpha scale},\\
\beta &=& \nu _{\perp} \left(2d - 6 + 3 \eta _K - 2 \eta _t
\right)/4 \label{beta scale},\\
\Omega &=& \nu_{\perp}\left(2-{3\over 2}\eta_K+\eta_t\right),
\end{eqnarray}
$\nu _z = \zeta\nu _{\perp} \label{nu z scale}$ and $\zeta = 2 -
{\eta _K \over 2}$. Note that $\alpha$ does {\it not} obey
hyperscaling, due to the strongly relevant disorder.

All of these exponents can be measured experimentally. The
specific heat can, of course, be measured by the usual
thermodynamic measurements. The spontaneous tilt angle $\theta_0$
can not be deduced from the position of the smectic $C$ Bragg
peak, since that peak is broad.

Fortunately, an alternative measure of $\theta_0$ can be deduced
from the dielectric or diamagnetic susceptibility tensors
$\chi_{ij}$ and $\epsilon_{ij}$. In the $A$ phase, one of the
principal axes of both tensors is along the applied field.  In the
$C$ phase, this axis rotates away from the applied field due to
the tipping  of the layers.  This rotation angle is proportional
to $\theta_0$.

The order parameter correlation lengths can be measured by light
scattering, which probes fluctuations in both the
dielectric($\epsilon_{ij}$) and diamagnetic($\chi_{ij}$)
susceptibility tensors. The full form of the light scattering is
extremely rich and complicated; we will defer a complete
description of it to a future publication. Here we will restrict
ourselves to pointing out that for $q_z=0$, the light scattering
intensity scales like $q_{\perp}^{2\eta_K-\eta_t-4}$ for
$q_{\perp}\gg\xi_{\perp}^{-1}$, is independent of $q_{\perp}$ for
$\left(\xi_{\perp}^{RF}\right)^{-1}\ll
q_{\perp}\ll\xi_{\perp}^{-1}$, and scales like $1\over q_{\perp}$
for $q_{\perp}\ll\left(\xi_{\perp}^{RF}\right)^{-1}$, where
$\xi_{\perp}^{RF}\propto \xi_{\perp}^{3-{3\over 2}\eta_K+\eta_t}$
as $T\to T_{AC}^+$. Thus light scattering data should easily allow
determination of $\xi_{\perp}(T)$ and the combination of exponents
$2\eta_K-\eta_t$. Fitting the T-dependence of $\xi_{\perp}$ to
$(T-T_{AC})^{-\nu_{\perp}}$ then determines $\nu_{\perp}$.

We now briefly sketch the derivation of our results. Our starting
point is an elastic energy for the layer displacement field $u$,
which is the {\it only} soft variable in the problem, since the
applied field locks the nematic director $\hat{n}$ \cite{LC}. For
the smectic $AC$ transition in a {\it pure} (i.e., disorder-free)
system, with an applied field freezing the director out, Grinstein
and Pelcovits \cite{GP} showed that the appropriate elastic energy
is:
\begin{eqnarray}
H_{pure} &=& \int d^dr  \left[{ K \over 2}(\nabla^2_{\perp}u)^2 +
{B \over  2}(\partial_zu)^2 - {g \over 2}(\partial_zu)
|\vec{\nabla}_{\perp}u|^2 \right.\nonumber\\&+&\left. {w \over 8}
\left|\vec{\nabla}_{\perp}u \right|^4 + {D_0(T) \over
2}|\vec{\nabla}_{\perp}u|^2 \right]. \label{2}
\end{eqnarray}

This model is very similar to that for a smectic $A$ in the {\it
absence} of an external field.  However,  because the rotational
symmetry is broken due to the external field, a new term $|\nabla_
{\perp}u|^2$, which hardens the directions orthogonal to
$\hat{z}$, is generated. Since $|\vec{\nabla}_{\perp} u|$ is
proportional to the tilt angle of the smectic layers, the
coefficient $D_0(T)$ is positive in the $A$ phase (favoring
alignment of the layer normal with the applied field), and
negative in the $C$ phase (favoring tilt of the layers). Hence, by
continuity, at $T=T_ {AC}$, $D_0(T)$ vanishes. In what follows, we
will assume that $D_0(T) \propto T-T_{AC}$ near $T_{AC}$.

The other terms in (\ref{2}) are simply those of the elastic
theory of a smectic $A$ in {\it zero} field, with one crucial
exception: in a smectic in {\it zero} field, rotation in variance
requires that $g=w=B$, while for the $AC$ in a {\it non-zero}
field problem, at all temperatures, even at $T = T_{AC}$, where $D
\rightarrow 0$ and softness is recovered, $g$ and $w$ are still
free, because rotation invariance is still broken.

To include the effects of the quenched disorder of the aerogel, we
add to the pure Hamiltonian (\ref{2}) random fields coupling to
$u$ and its gradients: giving us
\begin{eqnarray}
H=H_{pure}+\int d^dr
\left[\vec{h}\left(\vec{r}\right)\cdot\vec{\nabla}u +
V_p(u-\phi(\vec{r}))\right] \label{4}
\end{eqnarray}
where $\vec{h}(\vec{r})$ is a quenched random field that for
simplicity we take to be Gaussian zero distributed mean, and
characterized by short-ranged anisotropic correlations:
\begin{eqnarray}
\overline{h_i\left(\vec{r}\right)h_j\left(\vec{r}\,^{\prime}\right)} =
\left[\Delta_t\delta _{ij}^{\perp}+\Delta_c\delta_{ij}^z\right]\delta^d
\left(r-r^{\prime}\right).
\label{5}
\end{eqnarray}
The field $\phi(\vec{r})$ is also a quenched random field with
only short-ranged correlations, and is uniformly distributed
between 0 and $a$, the smectic layer spacing. The function
$V_p(u-\phi)$ is periodic with period $a$.

The physical interpretation of the quenched random fields
$\vec{h}(\vec{r})$ and $V_p(u-\phi)$ is very simple. Note that the
random field $\vec{h}$ incorporates random torques and random
compressions, coming from the $\perp$ and $z$ components of
$\vec{h}$, respectively. The function $V_p(u-\phi(\vec{r}))$
represents the tendency of the aerogel to pin the smectic layers
in a set of random positions $\phi(\vec{r})$, modulo the smectic
layer spacing $a$, which is why $V_p$ is periodic in its argument.

To compute self-averaging quantities, e.g., the disorder averaged
free energy, it is convenient to employ the replica ``trick'' that
relies on the identity $\overline{\log Z} = \lim\limits_{n
\rightarrow 0}{\overline{Z^n} - 1 \over n}$.  After replicating
and integrating over the disorder $\vec{h}(\vec{r})$ utilizing Eq.
\ref{5} we obtain \cite {foott}
\begin{eqnarray}
H[u_{\alpha}] &=& {1 \over 2} \int d^dr \sum^n_{\alpha = 1}
\left[K\left(\nabla_{\perp}^2u_{\alpha}\right)^2 + B\left(\partial
_zu_{\alpha}\right)^2\right.\nonumber\\&-&\left.g(\partial
_zu_{\alpha})|\vec{\nabla}_{\perp}u_{\alpha}|^2 + {w \over 4}
\left|\vec{\nabla}_{\perp}u_{\alpha}\right|^4\right.\nonumber\\&+&
\left.D_0(T)\left|\vec{\nabla}_{\perp}u_{\alpha}\right|^2
\right]\nonumber\\&-& {\Delta_t\over 2T}\int d^dr \sum^n_{\alpha,
\beta = 1} \vec{\nabla}_{\perp}u_{\alpha} \cdot
\vec{\nabla}_{\perp}u_{\beta} \label{6}
\end{eqnarray}
Assuming $D_0(T)$ is very small right at the phase transition the
noninteracting propagator $G_{\alpha \beta}\left(\vec{q} \right)
\equiv V^{-1}\left< u_{\alpha}(q) u_{\beta} (-q)\right>$ can be
easily obtained
\begin{eqnarray}
G_{\alpha \beta}(q) = TG \left(\vec{q} \right) \delta_{\alpha
\beta} + \Delta_t q^2_{\perp} G \left(\vec{q} \right)^2 \label{7}
\end{eqnarray}
with $G \left(\vec{q} \right) = 1/\left(Kq^4_{\perp} +
Bq^4_z\right)$.

We employ the standard momentum shell renormalization group (RG)
transformation.  The only novelty is that we will employ an
infinite hyper-cylindrical Brillouin zone:
$|\vec{q}_{\perp}|<\Lambda$, $-\infty<q_z<\infty$, where
$\Lambda\sim{1\over a}$ is an ultra-violet cutoff. We separate the
displacement field into high and low wave vector components
$u_{\alpha}\left(\vec{r}\right) = u_{\alpha}^<
\left(\vec{r}\right) +u_{\alpha}^> \left(\vec{r}\right)$, where
$u_{\alpha}^> \left(\vec{r}\right)$ has support in the
hyper-cylindrical shell $\Lambda e^{-\ell}<q_{\perp}<\Lambda$,
$-\infty<q_z<\infty$. We then integrate out the high wave vector
part $u_{\alpha}^> \left(\vec{r}\right)$, and rescale the length
and long wavelength part of the fields with $\vec{r}_{\perp} =
\vec{r}_{\perp}^{\prime} e^{\ell}$, $z =
z^{\prime}e^{\omega\ell}$, and $u_{\alpha}^< \left(\vec{r}\right)
= e^{\chi \ell}u^{\prime}_{\alpha}\left(\vec{r}\,
^{\prime}\right)$ so as to restore the $UV$ cutoff back to
$\Lambda$.

Evaluating the corrections $\delta B$, $\delta K$, $\delta
\Delta$, $\delta g$ and $\delta u$ and performing the rescalings
described above, we obtain the following $RG$ flow equations to
one loop order:
\begin{eqnarray}
{dB(\ell)  \over  d \ell} = \left(d -1-\omega+2\chi- {3 \over 16}
g_3\right)B, \label{10}
\end{eqnarray}
\begin{eqnarray}
{dK(\ell)  \over  d \ell} = \left(d -5+\omega+2\chi+ {1 \over 32}
g_3\right)K,\label{11}\\
{d\left(\Delta/T \right)(\ell)  \over  d \ell} = \left(d
-3+\omega+2\chi+ {1 \over 64} g_3 \right){\Delta\over T}, \label{12}\\
{dg(\ell)  \over  d \ell} = \left(d - 3 + 3\chi + {3 \over
32}g_3-{9\over 32}g_4\right)g, \label{15}
\end{eqnarray}
\begin{eqnarray}
{dD(\ell)  \over  d
\ell} &=& \left(d-3+\omega+2\chi+ {9 \over 16}g_3 - {5 \over
16}g_4 \right) D \nonumber\\&+&{5 \over 24}K
\left(g_4 - g_3\right), \label{13}
\end{eqnarray}
\begin{eqnarray}
{dw(\ell) \over  d \ell}&=& \left(d -5+\omega+4\chi-{3\over
32}{g_3^2\over g_4}\right)w\nonumber\\&+&\left({3\over
8}g_3-{15\over 32}g_4\right)w,\label{14}
\end{eqnarray}
where $g_2 \equiv \Delta \left(B/K^5 \right)^{1 \over
2}C_{d-1}\Lambda^{d-5}$, $g_3\equiv \left(g\over B\right)^2g_2$,
$g_4\equiv \left(w\over B\right)g_2$, $\epsilon = 5 - d$. These
$RG$ flow equations have two fixed points: one preserving rotation
invariance ($g_3= g_4$), which is unstable; and one with $g_3^* =
0$, $g_4^* = {32 \over 15} \epsilon$, which is stable and controls
the second-order phase transition. Analyzing the RG flows around
the stable fixed point in the standard way leads to the critical
properties, exponents, and scaling relations described earlier.

In summary, a theory of smectic $A-C$ phase transition in a field
in disordered media is developed.  We found the critical exponents
to first order in the $\epsilon = 5 - d$ expansion.  In addition,
we have made experimentally testable predictions for the
elasticity and fluctuations of this system in both phases, and at
the transition.

JT thanks the Aspen Center for Physics for their hospitality while
a portion of this work was being completed. LC thanks Yan Sheng
for her great support.


\begin{thebibliography}{99}
\bibitem{foott}
The random pinning $V_p(u-\phi)$ proves to be irrelevant at the
transition, and in the $C$ phase (though important in the $A$
phase). Since we are only discussing the transition here, we drop
it from (\ref 6).
\end{thebibliography}
\end{document}